\documentstyle[aps,prd]{revtex}

\begin{document}
\draft
\preprint{SNUTP-98-004}

\title{Loss of Quantum Coherence and Positivity of Energy Density
in Semiclassical Quantum Gravity}
\author{Sang Pyo Kim\footnote{Electronic address: sangkim@knusun1.kunsan.ac.kr}}
\address{Department of Physics, Kunsan National University,
Kunsan 573-701, Korea}
\author{Kwang-Sup Soh\footnote{Electronic address: kssoh@phya.snu.ac.kr} }
\address{Department of Physics Education, Seoul National
University, Seoul 151-742, Korea}

\maketitle
\begin{abstract}
In the semiclassical quantum gravity derived from the Wheeler-DeWitt equation,
the energy density of a matter field loses quantum coherence due to the
induced gauge potential from the parametric interaction with gravity
in a non-static spacetime. It is further shown that the energy density
takes only positive values and makes superposition principle hold true.
By studying a minimal massive scalar field in a FRW spacetime background,
we illustrate the positivity of energy density and obtain the classical
Hamiltonian of a complex field from the energy density in coherent states.
\end{abstract}

There has been a reviving interest in the back-reaction problem in
black hole and wormhole physics. The complete resolution of the
puzzle of back-reaction should be sought in quantum gravity, but at
present there is not known any viable theory of quantum
gravity, free from all problems. Though quantum gravity is not
available at hand, semiclassical treatment of a gravity-matter
system, quantized matter field and classical background spacetime,
sheds light on some important aspects of quantum effects. For instance,
the issue such as quantum interference or loss of quantum coherence
can be treated in semiclassical gravity without relying on quantum
gravity. In this semiclassical gravity there have
been developed two typical methods: one is the traditional approach
to semiclassical gravity \cite{birrel} and the other is the
so-called semiclassical quantum gravity approach \cite{kiefer}.
In the traditional approach, one first quantizes the matter field
on the fixed classical spacetime background, for instance,
a la the functional Schr\"{o}dinger equation. One then
employs diverse methods to evaluate the expectation value of
quantum stress-energy tensor and finally solves the semiclassical
Einstein equation, $G_{\mu \nu} = 8 \pi \langle \hat{T}_{\mu \nu}
\rangle$. One readily sees that quantum interference predominates
in the energy density expectation value and indeed leads to
possible negative energy density \cite{ford}.

On the other hand, in the semiclassical quantum gravity, one first
quantizes both the geometry and matter field within the framework
of canonical quantum gravity based on the Wheeler-DeWitt equation.
From the Wheeler-DeWitt equation in (semi-)classical regions one derives
the semiclassical quantum gravity: the Einstein-Hamilton-Jacobi equation
or equivalently the semiclassical Einstein equation, $G_{\mu
\nu} = 8 \pi \langle\langle \hat{T}_{\mu \nu} \rangle\rangle $
and the time-dependent functional Schr\"{o}dinger equation for the
matter field. There is one noticeable difference from the traditional approach:
the role of induced gauge potential \cite{brout,kim0,kim1,bertoni,kim2}.
The Wheeler-DeWitt equation for the gravity-matter system
is analogous to the Schr\"{o}dinger equation with zero energy for a
molecular system: the matter fields play the role of electrons
(fast particles) and the gravity that of  nuclei (slow particles).
In particular, it is observed that the off-diagonal elements of
the Hamiltonian and the induced gauge potential
cancel among themselves in the effective gravitational equation
in a matrix form \cite{kim0}. So one may expect the
expectation values of quantum energy density to differ
from each other in the two approaches to semiclassical gravity.

In this Rapid communication we study the effects of induced gauge
potential of the gravity-matter system on the energy density within
the framework of the semiclassical quantum gravity.
It is found that the energy density for a superposed quantum state
of matter field loses quantum coherence through the parametric interaction
with gravity in a non-static spacetime.
The loss of coherence in turn leads both to superposition principle
for any exclusive set of quantum states and to the positivity of energy density.
We compare these results with those from the traditional approach
in which superposition principle does not hold true due to quantum interference
among the quantum states and the energy density may take a negative value \cite{ford}.
For this purpose we elaborate further the formalism developed in Refs. \cite{kim1,kim2}
to make the role of gauge potential be exhibited for the superposed quantum state.
Through the study of a minimal massive scalar field in a FRW spacetime background,
we also illustrate how classical matter Hamiltonian emerges from
the decohered energy density.

Let us consider the Wheeler-DeWitt equation for a gravity-matter system
\begin{equation}
\Biggl[ - \frac{\hbar^2}{2 m_P^2} \nabla^2 - m_P^2 V(h_a)
+ {\bf H} \Bigl(\frac{\hbar}{i} \frac{\delta}{\delta \phi},
\phi; h_a \Bigr) \Biggr] \Psi(h_a, \phi) = 0,
\label{wdw eq}
\end{equation}
where $m_P = \frac{1}{\sqrt{G}}$ is the Planck mass,
$h_a$ and $\phi$ represent the superspace coordinates and
the matter field, respectively. The semiclassical quantum gravity
is obtained by applying the Born-Oppenheimer idea to separate the
Wheeler-DeWitt equation into an effective gravitational field (heavy particle) equation
and a time-dependent Schr\"{o}dinger equation for the matter field (light particle).
And then according to the de Broglie-Bohm interpretation the effective
gravitational field equation reduces to the Einstein-Hamilton-Jacobi equation
with quantum corrections. This scheme is valid in the (semi-)classical regions of
superspace where the gravitational wave function oscillates and therefrom decohered
(semi-)classical spacetime emerges.

We now wish to see how the matter field in the superposed quantum state
loses its quantum coherence through the parametric interaction with
gravity in a non-static spacetime. The matter field sector is
assumed to have a well-defined Hilbert space by whose bases the wave functions
can be expanded. We confine our attention to
a complex wave function and study the evolution of quantum field
along a single-branch of history, which includes the Vilenkin's
tunneling wave function but excludes the Hartle-Hawking's no-boundary
wave function \cite{hartle}. Then the wave function can always be written as
\begin{eqnarray}
\Psi (h_a, \phi) &=& \psi(h_a) \vert \Phi (\phi; h_a) \rangle,
\label{wave}
\\
\vert \Phi(\phi; h_a) \rangle &=& \sum_{n \in {\cal S}} c_n
\vert \Phi_n (\phi;h_a) \rangle,
\label{sup st}
\end{eqnarray}
where $\vert \Phi \rangle $ has a unit norm and $\{ \Phi_n \}$ forms
an orthonormal basis of the Hilbert space.
Any wave function that is superposed of more than two complex wave functions
of the Wheeler-DeWitt equation can be rewritten as Eq. (\ref{wave})
through complex transformations in the Hilbert space. For this
single-branch of history one is able to derive consistently the semiclassical
quantum gravity without many fundamental conceptual problems
mentioned in Ref. \cite{kiefer}.

The quantum state ({\ref{sup st}) depends on the superspace as parameters,
so a gauge potential is induced as the quantum state evolves on the
superspace. The induced gauge potential of the matter field is divided
into the diagonal and the off-diagonal part, ${\bf A}_D$ and ${\bf
A}_O$, respectively:
\begin{equation}
{\bf A} = \langle \Phi \vert i \hbar \nabla \vert \Phi \rangle
= \sum_{k, n \in {\cal S}} c^*_{k}c_{n} \langle \Phi_k
\vert i \hbar \nabla \vert \Phi_n \rangle
= {\bf A}_D + {\bf A}_O.
\end{equation}
In order to obtain correctly the semiclassical Einstein equation
with quantum back-reaction, it is necessary to treat the gauge
potential appropriately. According to Ref. \cite{kim0},
Eq. (\ref{wave}) can be written as ${\bf U}^T (\phi; h_a) \cdot {\Psi} (h_a)$,
where ${\bf U}$ is the column vector consisted of
$\vert \Phi_n \rangle$ and ${\bf \Psi}$ is the column vector consisted of $c_n \psi$.
There it has been shown that the off-diagonal elements of the
Hamiltonian and the induced gauge potential,
${\bf A}_O$, cancel among themselves in the effective gravitational equation
(18) of Ref. \cite{kim0}. The remaining energy density then consists of
only diagonal elements. We can show these facts more directly by appropriately
using the gauge potential but without relying on the matrix equation.
The idea is to multiply the gravitational wave function by a phase factor
from the diagonal part of the induced gauge potential and
to compensate it by multiplying the quantum state with the phase factor
of the opposite sign:
\begin{equation}
\Psi (h_a, \phi) =  \Biggl[e^{\frac{i}{\hbar} \int {\bf A}_{D, a} \cdot dh_a}
\psi (h_a)\Biggr] \times
\Biggl[e^{- \frac{i}{\hbar} \int {\bf A}_{D, a} \cdot dh_a} \vert \Phi
(\phi; h_a) \rangle \Biggr].
\end{equation}
The total wave function (\ref{wave}) is still invariant under the above gauge
choice. For the gravitational wave function of the form
\begin{equation}
\psi (h_a) = F(h_a) e^{\frac{i}{\hbar} S(h_a)},
\end{equation}
the de Broglie-Bohm interpretation separates the real and
imaginary parts of the Wheeler-DeWitt equation.
The semiclassical Einstein equation comes from the real part
\begin{eqnarray}
&&\Biggl[ \frac{1}{2m_P^2} \Bigl( \nabla S \Bigr)^2
- m_P^2 V + {\cal H}_D + {\cal H}_O + {\cal H}_{vac.}
- \frac{1}{m_P^2} {\bf A}_O \cdot \nabla S \Biggr]
\nonumber\\
&&{} + \Biggl[\frac{1}{2 m_P^2} {\bf A}_D^2
- \frac{1}{m_P^2} {\bf A} \cdot {\bf A}_D
- \frac{\hbar^2}{2m_P^2} \frac{\nabla^2 F}{F}
+ \frac{1}{2m_P^2} \langle \Phi \vert (i \hbar \nabla)^2
\vert \Phi \rangle  \Biggr] =0 ,
\label{sem ein}
\end{eqnarray}
where ${\cal H}_D$, ${\cal H}_O$, and ${\cal H}_{\rm vac.}$
are the diagonal, off-diagonal parts, and the
vacuum energy of the quantum back-reaction:
\begin{eqnarray}
\langle \Phi \vert {\bf H} \vert \Phi \rangle
&=& \sum_{k, n \in {\cal S}} c^*_{k}c_{n} \langle \Phi_k
\vert :{\bf H}: \vert \Phi_n \rangle
+ \sum_{n \in {\cal S}} c^*_{n} c_{n} \langle \Phi_n
\vert {\bf H} - :{\bf H}: \vert \Phi_n \rangle
\nonumber\\
&\equiv&  {\cal H}_D + {\cal H}_O + {\cal H}_{\rm vac.}.
\end{eqnarray}
The vacuum energy will be absorbed into the cosmological
constant and renormalize the potential, $V_{\rm ren.}$.
The imaginary part can be integrated for $F$ in terms of $S$
and put into Eq. (\ref{sem ein}) in a self-consistent way \cite{kim2}.

On the other hand, by subtracting Eq. (\ref{sem ein}) from Eq.
(\ref{wdw eq}) after acting by $\langle \Phi \vert$
and by introducing a cosmological (WKB) time
\begin{equation}
\frac{\delta}{\delta \tau} = \frac{1}{m_P^2} \nabla S \cdot \nabla ,
\end{equation}
one derives the time-dependent Schr\"{o}dinger equation for the matter
field
\begin{equation}
i \hbar \frac{\delta}{\delta \tau} \vert \Phi (\tau) \rangle =
\Biggl[{\bf H} + {\bf H}_{UQ} + {\bf H}_{NQ}
+ {\cal H}_{UC} + {\cal H}_{NC} \Biggr] \vert \Phi (\phi; h_a) \rangle,
\label{sch eq}
\end{equation}
where
\begin{eqnarray}
{\bf H}_{UQ} &=& - i \frac{\hbar}{m_P^2} {\bf A} \cdot \nabla
- \frac{\hbar^2}{2 m_P^2} \nabla^2 ,
\nonumber\\
{\bf H}_{NQ} &=& -\frac{\hbar^2}{m_P^2} \frac{\nabla F}{F} \cdot \nabla
\end{eqnarray}
are operator-valued quantum corrections such that
${\bf H}_{UQ}^{\dagger} = {\bf H}_{UQ}$,
${\bf H}_{NQ}^{\dagger} = - {\bf H}_{NQ}$, and
\begin{eqnarray}
{\cal H}_{UC} &=&  - {\cal H}_D - {\cal H}_O - {\cal H}_{\rm vac.}
+  \frac{1}{m_P^2} \nabla S \cdot {\bf A}
+ \frac{\hbar}{m_P^2} {\bf A} \cdot {\bf A}_D
- \frac{1}{2m_P^2} \langle \Phi \vert (i \hbar \nabla)^2
\vert \Phi \rangle,
\nonumber\\
{\cal H}_{NC} &=& - i \frac{\hbar}{m_P^2} {\bf A} \cdot \frac{\nabla
F}{F}
\end{eqnarray}
are $c$-numbers such that ${\cal H}_{UC}^* = {\cal H}_{UC}$,
${\cal H}_{NC}^* = - {\cal H}_{NC}$. Most of the $c$-numbers
contribute physically uninteresting phase factors and will not be
considered further. We briefly comment on the
unitarity of quantum field. There are two terms possibly violating the unitarity,
${\bf H}_{NQ}$ and ${\cal H}_{NC}$.  However, they are contributing the phase factors
\begin{equation}
\exp \Bigl( \frac{1}{i \hbar} \int \langle \Phi \vert {\bf H}_{NQ}
\vert \Phi \rangle \Bigr) = \exp \Bigl( \frac{1}{m_P^2}
\int \frac{ \nabla F}{F} \cdot {\bf A} \Bigr)
\end{equation}
and
\begin{equation}
\exp \Bigl(\frac{1}{i \hbar} \int {\cal H}_{NC} \Bigr)
= \exp \Bigl( - \frac{1}{m_P^2} \int \frac{ \nabla F}{F} \cdot {\bf A}
\Bigr).
\end{equation}
Therefore, they cancel each other, and Eq. (\ref{sch eq}) preserves
the unitarity as shown in Refs. \cite{bertoni,kim2}.

We shall now work with the semiclassical quantum gravity at the order of ${\cal O} (\hbar)$.
Recollecting that ${\bf A}$ is of the order of ${\cal O} (\hbar)$
and the terms in the second square bracket in Eq. (\ref{sem ein})
and ${\bf H}_{UQ}$, ${\bf H}_{NQ}$ in Eq. (\ref{sch eq}) are all of the order of
${\cal O} (\hbar^2)$, one obtains the equations for semiclassical quantum gravity
to the order of ${\cal O} (\hbar)$:
\begin{eqnarray}
&&\frac{1}{2m_P^2} \bigl( \nabla S_{(0)} \bigr)^2
- \frac{1}{m_P^2} {\bf A}_{(0),O} \cdot \nabla S_{(0)}
- m_P^2 V_{\rm ren.}  + {\cal H}_{(0),D} + {\cal H}_{(0),O} = 0,
\label{low gr eq}\\
&&i \hbar \frac{\delta}{\delta \tau} \vert \Phi_{(0)} (\tau) \rangle
= {\bf H} \vert \Phi_{(0)} (\tau) \rangle.
\label{low eq}
\end{eqnarray}
We make use of the well-known fact for a time-dependent quantum system
that when the basis of the exact quantum states of Eq. (\ref{low eq})
are chosen, the off-diagonal elements of the gauge potential are the
same as those of the Hamiltonian \cite{lewis}
\begin{equation}
\Bigl({\cal H}_{(0),O} \Bigr)_{kn} = \langle \Phi_{(0), k} \vert :{\bf
H}: \vert \Phi_{(0), n} \rangle
= \langle \Phi_{(0), k} \vert i \hbar \frac{\delta}{\delta \tau}
 \vert \Phi_{(0), n} \rangle
= \frac{1}{m_P^2} \Bigl( {\bf A}_{(0),O} \Bigr)_{kn} \cdot \nabla.
\end{equation}
So these off-diagonal elements cancel among themselves.
Equation (\ref{low gr eq}) becomes the time-time component of the
semiclassical Einstein equation in the form of Einstein-Hamilton-Jacobi equation
\begin{equation}
\frac{1}{2m_P^2} \Bigl(\nabla S_{(0)} \Bigr)^2
 - m_P^2 V_{\rm ren.}
+ \langle\langle \Phi_{(0)} \vert :{\bf H}: \vert
\Phi_{(0)} \rangle\rangle = 0,
\label{sqg}
\end{equation}
where ${\cal H}_{(0),D}$ is denoted by
\begin{equation}
\langle\langle \Phi_{(0)} \vert :{\bf H}: \vert \Phi_{(0)}
\rangle\rangle \equiv {\cal H}_{(0),D} = \sum_{n \in {\cal S}}
c^*_n c_n \langle \Phi_{(0), n} \vert :{\bf H}: \vert \Phi_{(0), n} \rangle.
\label{sqg ex}
\end{equation}
Note that for a positive definite ${\bf H}$ each term in Eq. (\ref{sqg})
takes positive value except for the trivial case of vacuum state. So the semiclassical
quantum gravity allows only the positive energy density.
Furthermore, for any two exclusive sets ${\cal S}_1$ and ${\cal S}_2$
such that ${\cal S}_1 \cap {\cal S}_2 =
\emptyset$, superposition principle, which applies to classical gravity,
also holds true for the quantum energy density:
\begin{equation}
\sum_{n \in {\cal S}_1 \cup {\cal S}_2}
 \langle\langle \Phi_{(0)}  \vert :{\bf H}: \vert
 \Phi_{(0)} \rangle\rangle = \sum_{n \in {\cal S}_1}
\langle\langle \Phi_{(0)} \vert :{\bf H}: \vert \Phi_{(0)}
\rangle\rangle + \sum_{n \in {\cal S}_2}
\langle\langle \Phi_{(0)} \vert :{\bf H}: \vert
\Phi_{(0)} \rangle\rangle.
\end{equation}
This is to be compared with that of the traditional approach
\begin{equation}
\langle \Phi_{(0)} \vert :{\bf H}: \vert \Phi_{(0)} \rangle
= \sum_{k, n \in {\cal S}} c^*_k c_n
\langle \Phi_{(0), k} \vert :{\bf H}: \vert \Phi_{(0), n} \rangle,
\label{con ex}
\end{equation}
where quantum interference among $k \neq n$ predominates in Eq. (\ref{con ex}).

In order to illustrate the formalism developed so far,
we shall consider a simple cosmological model with a minimal scalar field.
Let us consider the minimal massive scalar field in a
non-static Friedmann-Robertson-Walker universe with the metric
\begin{equation}
ds^2 = -N^2 dt^2 + a^2 (t) d \Omega_3^2.
\end{equation}
The corresponding Wheeler-DeWitt equation takes the form
\begin{equation}
\Biggl[\frac{2 \pi \hbar^2}{3 m_P^2 a} \frac{\partial^2}{\partial a^2}
- \frac{3m_P^2}{8 \pi} k a + \Lambda a^3 + \frac{1}{2 a^3}
\hat{\pi}_{\phi}^2 + \frac{m^2 a^3}{2}
\hat{\phi}^2 \Biggr] \Psi (a, \phi) = 0,
\label{frw eq}
\end{equation}
where $k = 1, 0, -1$ for a closed, flat and open universe, respectively,
and $\Lambda$ is the cosmological constant, and $\pi_{\phi} = a^3 \dot{\phi}$.
From Eq. (\ref{frw eq}) we derive the semiclassical Einstein equation
\begin{equation}
\Bigl(\frac{\dot{a}}{a} \Bigr)^2 + \frac{k}{a^2} +
\Lambda_{\rm ren.} = \frac{8 \pi}{3m_P^2 a^3}
\langle \langle \Phi_{(0)} (\tau) \vert :{\bf H}:
\vert \Phi_{(0)} (\tau) \rangle \rangle.
\end{equation}
The cosmological time $\frac{\partial}{\partial \tau}
= - \frac{4 \pi}{3m_P^2 a} \frac{\partial S(a)}{\partial a}
\frac{\partial}{\partial a}$ is identified with the comoving time
$t$ and no distinction will be made between $\tau$ and $t$.
The Hamiltonian for the scalar field can decomposed into
Fourier-modes\cite{birrel}:
\begin{equation}
{\bf H} = \sum_{\alpha} \Biggl[ \frac{1}{2 a^3} \hat{\pi}_{\alpha}^2 +
\frac{a^3}{2}  \Bigl(m^2 + \frac{\omega_{\alpha}^2}{a^2} \Bigr)
\hat{\phi}^2_{\alpha} \Biggr],
\end{equation}
where $\omega_{\alpha}^2$ denote the eigenvalues of $- \nabla^2$.
The exact quantum states for the Schr\"{o}dinger
equation are the number states, up to time-dependent phase
factors, constructed from the annihilation and creation operators \cite{kim2}
\begin{eqnarray}
\hat{b}_{\alpha} (\tau)  &=& i \Bigl(\varphi_{\alpha}^* (\tau)
\hat{\pi}_{\alpha} - a^3 (\tau) \dot{\varphi}^*_{\alpha} (\tau)
\hat{\phi}_{\alpha} \Bigr),
\nonumber\\
\hat{b}_{\alpha}^{\dagger} (\tau)  &=& - i \Bigl(\varphi_{\alpha} (\tau)
\hat{\pi}_{\alpha} - a^3 (\tau) \dot{\varphi}_{\alpha} (\tau)
\hat{\phi}_{\alpha} \Bigr),
\label{an-cr op}
\end{eqnarray}
where each mode satisfies the corresponding classical equation of motion
\begin{equation}
\ddot{\varphi}_{\alpha} + 3 \frac{\dot{a}}{a}
\dot{\varphi}_{\alpha} + \Bigl(m^2 +
\frac{\omega_{\alpha}^2}{a^2} \Bigr) \varphi_{\alpha} = 0.
\label{cl eq}
\end{equation}
These operators are chosen to satisfy the commutation relations
$[\hat{b}_{\alpha}, \hat{b}_{\beta}^{\dagger} ]
= \delta_{\alpha, \beta}$. Then the Hamiltonian has the oscillator
representation
\begin{equation}
{\bf H} = \sum_{\alpha} C_\alpha (\hat{b}_{\alpha}^{\dagger}
\hat{b}_{\alpha} + \hat{b}_{\alpha} \hat{b}_{\alpha}^{\dagger})
+ D_{\alpha} \hat{b}_{\alpha}^2 + D_{\alpha}^*
\hat{b}_{\alpha}^{\dagger 2},
\label{mat ham}
\end{equation}
where
\begin{eqnarray}
C_{\alpha} &=& \frac{\hbar^2 a^3}{2} \Biggl[\dot{\varphi}_{\alpha}^*
\dot{\varphi}_{\alpha} + \Bigl(m^2 + \frac{\omega_{\alpha}^2}{a^2}
\Bigr) \varphi_{\alpha}^* \varphi_{\alpha} \Biggr],
\nonumber\\
D_{\alpha} &=& \frac{\hbar^2 a^3}{2}
\Biggl[\dot{\varphi}_{\alpha}^2 + \Bigl(m^2 + \frac{\omega_{\alpha}^2}{a^2}
\Bigr) \varphi_{\alpha}^2 \Biggr].
\end{eqnarray}
Following Ref. \cite{kim2}, we obtain the induced gauge potential
\begin{equation}
{\bf A} = \sum_{\alpha} 2 C_\alpha \hat{b}_{\alpha}^{\dagger}
\hat{b}_{\alpha} + D_{\alpha} \hat{b}_{\alpha}^2 + D_{\alpha}^*
\hat{b}_{\alpha}^{\dagger 2}.
\label{mat gauge}
\end{equation}
One thus sees that the off-diagonal elements of the Hamiltonian in
Eq. (\ref{mat ham}) cancel exactly those of the gauge potential in Eq.
(\ref{mat gauge}). The Fock space constructed above can also be applied
to a minimal massless scalar field except for the zero-mode.

We now wish to show that quantum interference may lead to negative
energy density in the traditional approach and the loss of quantum
coherence always leads to positive energy density in the semiclassical
quantum gravity approach. First, let us consider the quantum state superposed of two
numbers states $\vert n_{\alpha}, \tau \rangle$ and
$\vert n_{\alpha}+2, \tau \rangle$. The reason for choosing these
quantum states is that particles are created or annihilated by pairs
for the minimal massive scalar field in the curved spacetime.
From the inequality $\Bigl(\frac{D_{\alpha}
+ D_{\alpha}^*}{C_{\alpha}} \Bigr)^2 < 4$,  one can find the unique state
\begin{equation}
\vert \Phi_{\alpha} (\tau) \rangle = \frac{1}{\sqrt{1
+ \epsilon^2_{\alpha}}} \Bigl[ \vert 0, \tau \rangle
+ \epsilon_{\alpha} \vert 2, \tau \rangle  \Bigr],
\end{equation}
that makes the normal ordered Hamiltonian
\begin{equation}
\langle \Phi_{\alpha} (\tau) \vert : {\bf H}_{\alpha} : \vert
\Phi_{\alpha} (\tau) \rangle = \frac{1}{1 + \epsilon^2_{\alpha}}
\Bigl[4 \epsilon_{\alpha}^2
C_{\alpha} + \sqrt{2} \epsilon_{\alpha}
(D_{\alpha} + D_{\alpha}^*) \Bigr],
\label{coh exp}
\end{equation}
have negative energy density when
\begin{equation}
0 > \epsilon_{\alpha} > - \frac{1}{2 \sqrt{2}}
\Bigl(\frac{D_{\alpha} + D_{\alpha}^*}{C_{\alpha}} \Bigr),
~~{\rm or} - \frac{1}{2 \sqrt{2}} \Bigl(\frac{D_{\alpha}
+ D_{\alpha}^*}{C_{\alpha}} \Bigr) > \epsilon_{\alpha} > 0,
\end{equation}
depending on the signs of $\Bigl(\frac{D_{\alpha}
+ D_{\alpha}^*}{C_{\alpha}} \Bigr)$.
We have thus shown that the minimal massive
scalar field can have the negative energy density \cite{ford}.
We compare this with the energy density in the semiclassical quantum gravity
\begin{equation}
\langle \langle \Phi_{\alpha} (\tau) \vert : {\bf H}_{\alpha}:
\vert \Phi_{\alpha} (\tau) \rangle \rangle = \frac{4 \epsilon_{\alpha}^2}{
1 + \epsilon^2_{\alpha}} C_{\alpha},
\label{dec exp}
\end{equation}
which is obviously positive definite.

Next, we show how classical matter Hamiltonian emerges
from the semiclassical quantum gravity. It is well-known that
coherent states of a quantum system have classical features.
Let us consider the coherent states for each mode
\begin{equation}
\vert v_{\alpha}, \tau \rangle = e^{ - \frac{| v_{\alpha} |^2}{2}}
\sum_{n_{\alpha} = 0}^{\infty}
\frac{v_{\alpha}^n}{\sqrt{n_{\alpha}!}}
\vert n_{\alpha}, \tau \rangle.
\end{equation}
It is straightforward to show
\begin{equation}
\langle \langle v_{\alpha} (\tau) \vert : {\bf H}_{\alpha} : \vert
v_{\alpha} (\tau) \rangle \rangle = \frac{\hbar^2 a^3}{2}
2(v_{\alpha}^* v_{\alpha})
\Biggl[\dot{\varphi}_{\alpha}^* \dot{\varphi}_{\alpha}
 + \Bigl(m^2 + \frac{\omega^2_{\alpha}}{a^2}
\Bigr){\varphi}_{\alpha}^* {\varphi}_{\alpha}  \Biggr].
\label{coh st}
\end{equation}
Note that Eq. (\ref{coh st}) is $2(v_{\alpha}^* v_{\alpha})$
times the vacuum expectation value.
Each ${\varphi}_{\alpha}$ satisfies the corresponding
classical equation of motion (\ref{cl eq}).
By defining classical modes
\begin{equation}
\varphi_{\alpha, c} = \hbar \sqrt{2(v_{\alpha}^* v_{\alpha})}
{\varphi}_{\alpha},
\end{equation}
we obtain the classical Hamiltonian of a complex field
$\pi_{\phi_c} = a^3 \dot{\phi_c}$:
\begin{equation}
\langle \langle v \vert : {\bf H} :
\vert v \rangle \rangle =
\frac{1}{2a^3} \pi_{\phi_c}^* \pi_{\phi_c}
+ \frac{m^2 a^3}{2} \phi_c^* \phi_c.
\end{equation}
The $\phi_c$ satisfies the same classical equation for the real scalar field
$\phi$.

In summary, within the context of the semiclassical quantum
gravity derived from the Wheeler-DeWitt equation, we have shown that superposition
principle holds true for any exclusive set of quantum states due to the loss of
quantum coherence and the energy density always takes positive value.
Furthermore, through the study
of a minimal massive scalar field in a non-static FRW spacetime it has been
proved that the loss of quantum coherence makes the energy density take positive
values only, which may take negative values
due to the quantum coherence in the traditional approach.
It should, however, be remarked that the induced gauge potential
vanishes for a static spacetime and the coherence of the matter field
recovers even in the semiclassical quantum gravity.
The loss of quantum coherence (decoherence) is not due to the
interaction of the matter field with an environment but entirely due to
the interaction of the matter field with the gravity.
We have also recovered the classical Hamiltonian of a complex field
from the coherent state in the semiclassical quantum gravity.
The result would have some cosmological
implications, since any matter field in the Universe that changes
the spacetime significantly should always provide positive energy density
but certain gravity phenomena require negative energy density.
It would also be interesting to test any possible modification to the quantum physics
in the laboratory scale due to the ontological influence of the quantum cosmology.

This work was supported by the BSRI Program under
Project No. BSRI-97-2418, BSRI-97-2427 and by the Center for Theoretical
Physics, Seoul National University.


\begin{references}
\bibitem{birrel} For a review and references, see
N. D. Birrel and P. C. W. Davies,
{\it Quantum Fields in Curved Spacetime} (Cambridge University press,
Cambridge, 1982).
\bibitem{kiefer} For a review and references, see
C. Kiefer, in {\it Canonical Gravity: From Classical to Quantum}, edited by J. Ehlers
and H. Friedrich (Springer, Berlin, 1994); and for the conceptual
problems of the semiclassical quantum gravity, see K. V.
Kuchar, in {\it 4th Canadian Conference on
General Relativity and Relativistic Astrophysics}, edited by
G. Kunstatter, D. E. Vincent, and J. G. Williams (World Scientific,
Singapore, 1994).
\bibitem{ford} H. Epstein, V. Glaser, and A. Jaffe,
Nuovo Cimento {\bf 36}, 1016 (1965);
L. H. Ford, Proc. R. Soc. Lond. A {\bf 364}, 227 (1978);
C.-I Kuo and L. H. Ford, Phys. Rev. D, {\bf 47}, 4510;
C.-I. Kuo, Ph. D. Thesis, Tufts University, 1994 (unpublished);
Il Nouvo Cimento {\bf 112B}, 629 (1997); M. J. Pfenning, Ph. D. Thesis, Tufts
University, gr-qc/9805037 (1998).

\bibitem{brout} R. Brout, Found. Phys. {\bf 17}, 603 (1987);
R. Brout, G. Horwitz, and D. Weil, Phys. Lett. {\bf 192B}, 318 (1987);
R. Brout, Z. Phys. B {\bf 68}, 339 (1987);
T. P. Singh and T. Padmanabhan, Ann. Phy. (N.Y.)
{\bf 196}, 296 (1989);
R. Brout and G. Venturi, Phys. Rev. D {\bf 39}, 2436 (1989);
J. Kowalski-Glikman and J. C. Vink, Class.
Quantum Grav. {\bf 7}, 901 (1990);
R. Balbinot, A. Barletta, and G. Venturi,
Phys. Rev. D {\bf 41}, 1848 (1990);
D. P. Datta, Mod. Phys. Lett. A {\bf 8}, 191 (1993);
C. Gundlach, Phys. Rev. D {\bf 48}, 1700 (1993).
\bibitem{kim0} S. P. Kim and S.-W. Kim, Phys. Rev. D {\bf 49},
R1679 (1994).
\bibitem{kim1} S. P. Kim, Phys. Rev. D {\bf 52}, 3382 (1995);
S. P. Kim, Phys. Lett. {\bf 205A}, 359 (1995).
\bibitem{bertoni} C. Bertoni, F. Finelli, and G. Venturi,
Class. Quantum Grav. {\bf 13}, 2375 (1996).
\bibitem{kim2} S. P. Kim, Phys. Rev D {\bf 55}, 7511 (1997); Phys.
Lett. {\bf 236A}, 11 (1997).
\bibitem{hartle} A. Vilenkin, Phys. Lett. {\bf 117B}, 25 (1982);
J. B. Hartle and S. W. Hawking, Phys. Rev. D {\bf 28}, 2960 (1983).
\bibitem{lewis} H. R. Lewis, Jr. and W. B. Riesenfeld,
J. Math. Phys. {\bf 10}, 1458 (1969).

\end{references}
\end{document}